\newcommand{\BE}{\begin{eqnarray}}
\newcommand{\EE}{\end{eqnarray}}
\newcommand{\be}{\begin{eqnarray}}
\newcommand{\ee}{\end{eqnarray}}
\newcommand{\BEN}{\begin{eqnarray*}}
\newcommand{\EEN}{\end{eqnarray*}}
\newcommand{\ben}{\begin{eqnarray*}}
\newcommand{\een}{\end{eqnarray*}}
\newcommand{\BA}{\begin{array}}
\newcommand{\EA}{\end{array}}
\newcommand{\BLN}{\begin{align*}}
\newcommand{\ELN}{\end{align*}}

\newcommand{\cd}{\cdot}
\newcommand{\f}{\frac}

\newcommand{\bl}{\begin{lstlisting}}
\newcommand{\el}{\end{lstlisting}}

\newcommand\disp{\displaystyle}

\newcommand{\Ai}{\operatorname{Ai}}
\newcommand{\Bi}{\operatorname{Bi}}
\newcommand{\Li}{\operatorname{Li}}

\newcommand{\op}[1]{\operatorname{#1}}
\newcommand{\Ord}{\mathcal{O}}
\newcommand{\noi}{\noindent}

\documentclass[12pt,a4paper,twoside]{article}

\usepackage{enumerate}
\usepackage{cite}
\usepackage[english]{babel}
\usepackage[latin1]{inputenc}
\usepackage{subfigure}
\usepackage[T1]{fontenc}
\usepackage{graphicx}
\usepackage{latexsym}
\usepackage{amsmath,amssymb,mathrsfs}
\usepackage{ifpdf}
\usepackage{cancel} 
\usepackage{amsmath, amsfonts, amsthm, amssymb, amscd}
\usepackage{graphicx}
\usepackage{makeidx}
\usepackage{sectsty}

\usepackage[format=hang,margin=1cm,textfont=small,labelfont=up]{caption}[2013/02/03] 
\usepackage{eufrak}
\usepackage{fancyhdr}
\usepackage[colorlinks=false,pdfborder={0 0 0}]{hyperref}

\sectionfont{\large}
\subsectionfont{\normalsize}

\newtheoremstyle{theorems}  
{\topsep}   
{\topsep}   
{\itshape}  
{10pt}      
{\bfseries} 
{.}         
{5pt} 	  
{}          

\newtheoremstyle{definitions}  
{\topsep}   
{\topsep}   
{}          
{10pt}      
{\bfseries} 
{.}         
{5pt} 	  
{}          

\theoremstyle{theorems}

\begin{document}
	
	\pagestyle{myheadings}
	\fancyhf{}
	\fancyhead[L]{}
	\fancyhead[R]{}
	
	\title{\Large{Area-width scaling in generalised Motzkin paths}}

	\author{\normalsize{Nils Haug\footnote{School of Mathematical Sciences, Queen Mary University of London, London, E1 4NS, United Kingdom}, Thomas Prellberg$^*$ and Grzegorz Siudem\footnote{Faculty of Physics, Warsaw University of Technology, Koszykowa 75, PL-00-662 Warsaw, Poland}}}
	
	\date{\normalsize{\today}}
	
	\maketitle
	
	\begin{abstract}

\footnotesize{\noindent We consider a generalised version of Motzkin paths, where horizontal steps have length $\ell$, with $\ell$ being a fixed positive integer. We first give the general functional equation for the area-width generating function of this model. Using a heuristic ansatz, we then derive the area-width scaling behaviour in terms of a scaling function in one variable for the special cases of Dyck, (standard) Motzkin and Schr\"oder paths, before generalising our approach to arbitrary $\ell$. We then rigorously derive the tricritical scaling of Schr\"oder paths by applying the generalised method of steepest descents to the known exact solution for their area-width generating function. Our results show that for Dyck and Schr\"oder paths, the heuristic scaling ansatz reproduces the rigorous results.}

\end{abstract}

	\sectionfont{\normalsize}
	\subsectionfont{\normalsize}
	
\section{Introduction}
\label{introduction}

Vesicles consist of a closed lipid membrane enclosing a fluid and act as containers for molecules inside of biological cells \cite{Alberts07}. Depending on parameters such as temperature and the osmotic pressure acting on the outside of the membrane, they favour different conformations. In the case of a high external pressure, the vesicles tend to minimise their volume, whereas more spatially extended configurations are typical in the case of low pressure.

Two-dimensional vesicles can be modelled as self-avoiding polygons (SAP), either in the continuum or on the lattice \cite{Guttmann12,Leibler89,Fisher91}. In this case, the volume of the vesicle becomes the area of the polygon and the perimeter is the distance covered when travelling around the polygon once. The partition function of vesicles, modelled as SAP on $\mathbb{Z}^2$ with fixed perimeter $m$, zero bending rigidity and subject to an external pressure $\epsilon$, is given by
\be 
Z_m(q) = \sum_{n=0}^\infty p_{m,n} q^n,
\ee
where $p_{m,n}$ is the number of SAP of perimeter $m$ and area $n$ and $q=\exp(-\epsilon)$.

\noi The corresponding area-perimeter generating function is defined as
\be 
P(t,q) = \sum_{m=0}^\infty Z_m(q) t^m = \sum_{m=0}^\infty \sum_{n=0}^\infty p_{m,n} t^m q^n,
\label{eq:SAP_gf}
\ee
and can be interpreted physically as the grand-canonical partition function of the model, where both area and perimeter of the vesicle can fluctuate. The shape of the phase diagram of this ensemble has been discussed in \cite{Fisher91}. In particular, it was shown that the radius of convergence $t_c(q)$ of $P(t,q)$, seen as a series in $t$, is positive for $q\leq 1$ and zero for $q>1$. The point $(t,q)=(t_c(1),1)$ is called a tri-critical point \cite{Lawrie84}.

Based on exact enumeration data, Richard, Guttmann and Jensen conjectured in \cite{Richard01} that the generating function of \emph{rooted} SAP, given by $P_{r}(t,q) = t \f{d}{dt} P(t,q)$ satisfies a $q$-functional equation of finite degree with polynomial coefficients. Subject to this conjecture, they argued that in a region around the tri-critical point, the singular part of $P_r(t,q)$ should obey the scaling relation
%
\be 
{P_r}^{(\text{sing})}(t_c(1)-z\epsilon^\phi,1-\epsilon) = \epsilon^{\theta} F\left(z\right)+o(\epsilon^{\theta}),
\label{eq:scaling_behaviour_SAP}
\ee
where the critical exponents are $\theta=1/3$ and $\phi=2/3$, and 
\be 
F(z) = b_0 \f{\op{Ai}'(b_1 z)}{\op{Ai}(b_1 z)}.
\label{eq:general_Airy_scaling_function}
\ee
Here, $\op{Ai}(z)$ is the \href{http://dlmf.nist.gov/9}{Airy function} \cite{NIST}, defined for $z \in \mathbb{C}$ and $c_{\pm}=e^{\pm i\pi/3}$ as
\be 
\Ai(z)=\f{1}{2\pi i}\int_{c_-\infty}^{c_+\infty} \exp\left(\f{u^3}{3}-zu\right)du,
\label{eq:def_Airy_function}
\ee
and $b_0$ and $b_1$ are constants.

From the scaling function, interesting statistical properties of the model, such as the distribution of areas in the infinite perimeter ensemble, can then be deduced \cite{Richard09}. 

Despite extensive research however, there are almost no rigorous results concerning the enumeration of SAP \cite{Guttmann12}. Therefore, a mathematical validation of the conjecture made in \cite{Richard01} appears to be currently out of reach. 

Progress in understanding the problem can be made by considering directed subclasses of SAP, for which a functional equation for the area-peri\-meter generating function is known. Examples for such models are Dyck paths, staircase polygons and directed column-convex polygons. From the functional equation, it is possible to extract the tri-critical scaling behaviour of the model, either by using a heuristic scaling ansatz \cite{Richard02} or by carrying out rigorous saddle point analysis on the exact solution for the generating function. The latter method has been applied in \cite{Prellberg95,Haug15} to the exact solutions for the generating functions of staircase polygons and Dyck paths, which are known from \cite{Brak90,Flajolet80}. For both models, it was shown that the scaling function is given by Eq.\eqref{eq:general_Airy_scaling_function}, with model-dependent values of $b_0$ and $b_1$.

It is natural to try to generalise the scaling result for Dyck paths to other one-dimen\-sional directed lattice walks, such as Motzkin and Schr\"oder paths \cite{Krattenthaler15}. In this case, it is easier to consider the width, i.e. the distance between the start and the end point of the path rather than the perimeter. For Motzkin paths, which are closely related to RSOS configurations, the solution for the area-width generating function has been derived in \cite{Owczarek09}, and for Schr\"oder paths, the area-width generating function was derived in \cite{Oppenheim02}. However, no corresponding scaling forms have been extracted yet. 

Out of this motivation, we here analyse the scaling behaviour of a model called $\ell$-Motzkin paths, with steps $(1,1)$, $(1,-1)$ and $(\ell,0)$, where $\ell$ is a fixed positive integer. This model has been studied previously in the combinatorics literature with a focus on bijections  \cite{Pergola02}. Motzkin and Sch\"oder paths are included in this family of walks and correspond to the cases $\ell=1$ and $2$, respectively, whereas Dyck paths can be identified with the limiting case $\ell=\infty$ \cite{Haglund15}.

After defining the model of $\ell$-Motzkin paths, we will re-derive the scaling behaviour for Dyck paths known from \cite{Haug15} by heuristically inserting a single-variable scaling ansatz into the functional equation for the generating function. This approach will then be extended to the cases of (standard) Motzkin and Schr\"oder paths, the scaling behaviour of which has not been analysed yet. Then we will generalise our results to $\ell$-Motzkin paths with arbitrary $\ell$.

As a further result, we give an alternative derivation of the exact solution for the area-width generating function of Schr\"oder paths, which can be expressed in terms of a quotient of basic hypergeometric series, similar to the well-known expression for Dyck paths \cite{Flajolet80}. From this we obtain the associated scaling form by rigorous saddle point analysis.

Our results show that the heuristic scaling ansatz reproduces the rigorous results for Dyck and Schr\"oder paths. Moreover, we obtain the same scaling form for all values of $\ell$, and therefore in particular for Motzkin paths.

\noi We will begin by precisely defining the model we consider.

\section{The model of $\ell$-Motzkin paths}
\label{model}

\begin{figure}[htbp]
\centering
\includegraphics[width=0.6\textwidth]{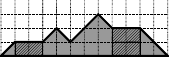}
\caption{A Schr\"oder path of width 12, with two horizontal steps of length two, three (2$\times 1$)-rectangles below these steps (hatched), four pairs of up/down steps, area 12 below these steps, and thus total area 18.}
  \label{fig:schroederpath}
\end{figure}

Given $\ell\in\mathbb{N}$ and $s\in \mathbb{Z}_{\geq 0}$, we define an $\ell$-Motzkin path of $s$ steps to be a lattice walk $(x_i,y_i)_{i=0}^s$ on $\mathbb{Z}_{\geq 0}^2$ such that $(x_0,y_0)=(0,0)$ and from any point $(x,y)$ on the path, the walker can either step towards $(x+1,y+1)$, $(x+1,y-1)$ or towards $(x+\ell,y)$, corresponding to an up-, down- or horizontal step, respectively. Moreover, the path needs to end on the horizontal line, i.e. $(x_s,y_s)=(m,0)$, where $m$ is the \emph{width} of the $\ell$-Motzkin path. Fig.\,\ref{fig:schroederpath} shows an example trajectory for the case $\ell=2$. Since we will only consider $\ell$-Motzkin paths in this paper, we will shortly refer to them as $\ell$-paths from now on.

\noi For given $\ell \in \mathbb{N}$, we define the generating function
\be 
G(s,u,p,q) = \sum_{k=0}^\infty\sum_{l=0}^\infty\sum_{v=0}^\infty\sum_{w=0}^\infty c_{k,l,v,w}\,s^k\,u^l\,p^v\,q^w,
\label{eq:def_general_gf}
\ee
where $c_{k,l,v,w}$ is the number of paths with $k$ horizontal steps, $\f{l}{2}$ pairs of up- and down-steps, $v$ ($\ell\times1$)-rectangles under all the horizontal steps, and $w$ unit squares under all the up- and down-steps (including the half unit squares directly underneath these steps). Thus the weight $u$ is associated to the total number of up- and down-steps, $q$ corresponds to the area under these steps, measured in unit squares of the lattice, and $s$ and $p$ weight the number of horizontal steps and the number of $(\ell\times 1$)-rectangles underneath these steps, respectively. For example, the weight of the trajectory shown in Fig. \ref{fig:schroederpath} is $s^2u^4 p^3 q^{12}$. Note that  there is no explicit $\ell$-dependence in $G(s,u,p,q)$. Instead, the area-width generating functions for $\ell$-paths of different $\ell$ are obtained by choosing appropriate values for $s, u$ and $p$, as will be explained below.

\begin{figure}[htbp]
\centering
\includegraphics[width=0.65\textwidth]{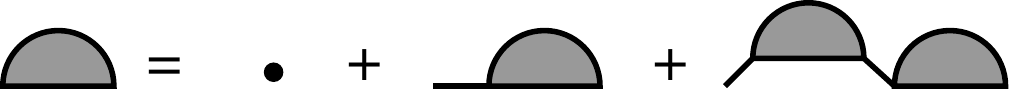}
\caption{Graphical interpretation of Eq.\eqref{eq:general_functional_eq}. An $\ell$-path either consists of zero steps, or it starts with a horizontal step, followed by an $\ell$-path, or it starts with an up-step, followed by an $\ell$-path, followed by a down-step, followed by another $\ell$-path.}
  \label{fig:functional_equation}
\end{figure}

A functional equation for $G(s,u,p,q)$ can be obtained by noting that for a given $\ell$, the set of all $\ell$-paths can be divided into the following three subsets. In Fig. \,\ref{fig:functional_equation} we  graphically illustrate this decomposition. The first subset only contains the trajectory of zero steps, which has weight 1. The second subset consists of all paths which start with a horizontal step, followed by a path (possibly of zero steps). The weight of a path in this set is thus the weight $s$ of the horizontal step at the beginning, times the weight of the path attached to this initial step. Finally, the third subset contains all the $\ell$-paths which start with an up-step. Their weight is  given by the weight $u$ of this initial up-step times the weight $u$ of its complimentary down-step, times the weight of the path in between these two steps, times the weight of the path following the down-step. Moreover, the two triangular regions below the initial up-step and the corresponding down-step together contribute one unit of area to the total area underneath the diagonal steps, which is accounted for by a factor of $q$. Since the path between the initial up-step and its complimentary down-step is elevated by one, each up- or down-step in between generates one further area of unit size, weighted by $q$, and each horizontal step generates an $(\ell\times 1)$-rectangle, weighted by $p$. Summing over the weights of the paths in all three subsets, this leads to the functional equation
\be 
G(s,u,p,q) = 1 + s\,G(s,u,p,q) + qu^2G(ps,qu,p,q)\,G(s,u,p,q).
\label{eq:general_functional_eq}
\ee
Note that by iteration of Eq.\eqref{eq:general_functional_eq}, we obtain the continued fraction representation
\be 
G(s,u,p,q) = \f{1}{1-s-\disp\f{qu^2}{1-ps-\disp\f{q^3u^2}{1-p^2s-\disp\f{q^5u^2}{1-p^3s-\dots}}}},
\ee
which can be used to approximate $G(s,u,p,q)$ numerically.

\noi In the following, we will consider the generating function
\be 
G^{(\ell)}(a,t,q) = \sum_{k=0}^\infty\sum_{l=0}^\infty\sum_{m=0}^\infty p_{k,m,n} a^k t^m q^n,
\label{eq:area-step_gf}
\ee
where $p_{k,m,n}$ is the number of paths with $2k$ diagonal steps, width $m$ and total area $n$ enclosed between the trajectory and the bottom line, counted in units of lattice cells. We will refer to $G^{(\ell)}(a,t,q)$ as the area-width generating function. Since each horizontal step of an $\ell$-path increases the width of the path by $\ell$ and each $(\ell\times 1)$-rectangle increases the total area, measured in units of lattice cells, by $\ell$, we have the identity
\be 
G^{(\ell)}(a,t,q) = G(t^\ell,\sqrt a t,q^\ell,q).
\label{eq:corresp_area_width_gf_general_gf}
\ee
\noi Substituting Eq.\eqref{eq:corresp_area_width_gf_general_gf} into Eq.\eqref{eq:general_functional_eq}, we obtain the functional equation
\be 
G^{(\ell)}(a,t,q) = 1 + t^\ell G^{(\ell)}(a,t,q) + aqt^2 G^{(\ell)}(a,qt,q)G^{(\ell)}(a,t,q).
\label{eq:general_funeq_area-length}
\ee
\noi For $q=1$, Eq.\eqref{eq:general_funeq_area-length} is solved by
\be 
G^{(\ell)}(a,t,1) = \f{1-t^\ell-\sqrt{(1-t^\ell)^2-4at^2}}{2at^2},
\label{eq:sol_gen_funeq_q=1}
\ee
and setting $a=1$ in Eq.\eqref{eq:sol_gen_funeq_q=1}, we obtain the generating functions of the Motzkin numbers for $\ell=1$ and the large Schr\"oder numbers for $\ell=2$ (\hspace{-0.125mm}\cite{wolfram_schroeder,wolfram_motzkin} and \href{https://oeis.org/A001006}{A001006} and \href{https://oeis.org/A006318}{A006318} in \cite{OEIS}).

For given $\ell \in \mathbb{N}$ and real $a>0$, we denote the smallest positive value for which the discriminant $(1-t^\ell)^2-4at^2$ vanishes by $t_c$ and define $G_c=G^{(\ell)}(a,t_c,1)$. From Eq.\eqref{eq:sol_gen_funeq_q=1} it follows that
\be 
G_c = \f{1-t_c^{\ell}}{2 a t_c^2}=\f1{\sqrt a t_c}.
\label{eq:Gc_general}
\ee
If $|t|<1$ and we let $\ell$ tend to infinity, then the weight $t^\ell$ associated to horizontal steps becomes zero, thus $G^{(\infty)}(a,t,q)=G(0,\sqrt{a}t,0,q)$ satisfies the functional equation
\be 
G^{(\infty)}(a,t,q) = 1 + a q t^2 G^{(\infty)}(a,qt,q)G^{(\infty)}(a,t,q).
\label{eq:funeq_DP}
\ee
In this case, the parameter $t$ only appears in powers of the product $at^2$ and therefore $a$ can be set equal to one without loss of generality. We write $G^{(\infty)}(t,q)\equiv G^{(\infty)}(1,t,q)$. Eq.\eqref{eq:funeq_DP} is then readily identified as the functional equation for the area-width generating function of Dyck paths \cite{Flajolet80}. If $q=1$, it is solved by the generating function of the Catalan numbers (\hspace{-0.125mm}\cite{Aigner07} and \href{https://oeis.org/A000108}{A000108} in \cite{OEIS}), and for general $q$, the solution was given in \cite{Flajolet80}. 

In the next section, we are now going to analyse the scaling behaviour of $G^{(\ell)}(a,t,q)$ around the point $(a,t,q)=(a,t_c,1)$ by using a heuristic ansatz. We will begin by treating the case of Dyck paths ($\ell=\infty$), for which the scaling function has been extracted rigorously via the method of steepest descents in \cite{Haug15}. Then we will apply the same approach to Motzkin and Schr\"oder paths ($\ell=1$ and $\ell=2)$, for which no scaling form has been derived yet in the literature, before generalising our approach to arbitrary $\ell$.

	\section{Heuristic scaling ansatz}
	\label{section:scaling_ansatz}

Given the known scaling behaviour of Dyck paths, the heuristic approach consists of assuming that in the vicinity of the point $(a,t,q)=(a,t_c,1)$, also the area-width generating functions of other $\ell$-paths satisfy a similar scaling relation. More precisely, we expect that there is a value $z_-<0$ such that for $z \in (z_-,\infty)$ and $\epsilon\to 0^+$,
\be 
G^{(\ell)}\big(a,t(z,\epsilon),1-\epsilon\big) = G_c+\epsilon^\theta F_0(a,z)+\epsilon^{2\theta}F_1(a,z)+\mathcal{O}(\epsilon^{3\theta})
\label{eq:scaling_behaviour}
\ee
where $t(z,\epsilon)=t_c-z\,\epsilon^{\phi}$; $\theta$ and $\phi$ are positive, non-integer critical exponents, and $F_0(a,z)$ and $F_1(a,z)$ are analytic functions for $z\in (z_-,\infty)$. The function $F_0(a,z)$ is the scaling function.

The heuristic approach is non-rigorous since it makes the assumption that $G^{(\ell)}\big(a,t(z,\epsilon),1-\epsilon\big)$ admits an expansion of the form \eqref{eq:scaling_behaviour}. Note, however that the method of dominant balance used below forms part of a rigorous method for deriving the area limit distribution of two-dimensional polygon models \cite{Richard09}.

For better readability, we omit the dependence of $a$ from now on and write $F_0(a,z)\equiv F(z)$. We define 
\be 
G^{(\ell)}_{sc}(a,z,\epsilon) = G_c+\epsilon^\theta F(z)+\epsilon^{2\theta}F_1(z).
\label{eq:def_Gscaling}
\ee

By setting $z=\tau \epsilon^{-\phi}$ with $0<\tau<t_c$, it follows from the positivity of the coefficients of the generating function $G^{(\ell)}(a,t,z)$ that for $a>0$,
\be
\lim_{z\to\infty}F(z)=-\infty.
\label{eq:limit_behaviour_F}
\ee

Following \cite{Richard02}, we now insert the RHS of Eq.\eqref{eq:scaling_behaviour} into the functional equation \eqref{eq:general_funeq_area-length}. Using a dominant balance argument, this uniquely determines the values of $\theta$ and $\phi$ and leads to an ODE for the function $F(z)$.

We will begin by validating the heuristic approach by reproducing the result for the scaling function of Dyck paths, known from \cite{Haug15}.

\subsection{Dyck paths ($\ell=\infty$)}
	
The area-width generating function of Dyck paths satisfies Eq.\eqref{eq:funeq_DP}, where, as explained above, $a$ can be set to one without loss of generality. Substituting $a=1$ and $t^\ell=0$ into the solution for $q=1$ given in Eq.\eqref{eq:sol_gen_funeq_q=1}, we obtain the critical values
	\begin{equation}
	t_c=\f{1}{2} \text{ and } G_c=2.
	\label{eq:critical_vals_dp}
	\end{equation}
Now we define the function
\begin{align*}
\Phi_\infty(z,\epsilon)=&1-G^{(\infty)}_{sc}(z,\epsilon)+(1-\epsilon)\,t(z,\epsilon)^2\,G_{sc}^{(\infty)}(z+t_c\epsilon^{1-\phi}-z\epsilon,\epsilon)\,G_{sc}^{(\infty)}(z,\epsilon),
\label{eq:Phi_DP}
\end{align*}
where $G_{sc}^{(\infty)}(z,\epsilon)\equiv G_{sc}^{(\infty)}(1,z,\epsilon)$ is given by Eq.\eqref{eq:def_Gscaling}, with an unknown function $F(z)$. Under the assumption that Eq.\eqref{eq:scaling_behaviour} holds, it follows from Eq.\eqref{eq:funeq_DP} that 
\be
\Phi_\infty(z,\epsilon)=\mathcal{O}(\epsilon^{3\theta}) \quad(\epsilon \to 0^+).
\label{eq:correction_to_scaling}
\ee
Expanding  $\Phi_\infty(z,\epsilon)$  into a series in  $\epsilon$, we obtain
\begin{align*}
\Phi_\infty(z,\epsilon)~=~& (1-G_c+t_c^2G_c^2)\\
~+~&\epsilon^\theta\left( (2t_c^2G_c-1)F(z)\right)\\
~+~&\epsilon^{2\theta}\left((2t_c^2G_c-1)F_1(z)+t_c^2F(z)^2\right)\\
~+~&\epsilon^{\phi}\left(-2t_cG_c^2 z\right)\\
~+~&\epsilon^{1-\phi+\theta}(t_c^3G_cF'(z))\\
~+~&\epsilon^{1-\phi+2\theta}(t_c^3(F(z)F'(z)+G_cF_1'(z)))\\
~+~&\epsilon^{\phi+\theta}(-4zt_cG_cF(z))\\
~+~&\mathcal{O}(\epsilon^{3\theta}).
\end{align*}
The constant coefficient and the one of order $\epsilon^\theta$ are zero by virtue of Eq.\eqref{eq:critical_vals_dp}. For Eq.\eqref{eq:correction_to_scaling} to hold, the coefficient of the order of $\epsilon^{2\theta}$ in the above equation needs to be cancelled by another coefficient, hence one of the other exponents needs to equal $2\theta$. If $2\theta=\phi+\theta$, thus $\theta=\phi$, then the term of order $\epsilon^\phi$ in the above equation could not be cancelled by any other term unless $\theta=1$, which is impossible by the assumption that $\theta$ is not an integer. Likewise, it is impossible that $2\theta=1-\phi+2\theta$, since $\phi$ is assumed to be non-integer. The third possibility is that $2\theta=\phi$, in which case the only way to obtain a solution $F(z)$ analytic at zero is to also have $2\theta=1-\phi+\theta$. The critical exponents hence necessarily satisfy the equations
		\begin{equation*}
		2\theta-\phi=0\text{ and }
		\theta+\phi=1,
		\end{equation*}
and thus $\theta=1/3$ and $\phi=2/3$. Inserting these exponents, the above equation simplifies to
\begin{equation*}
\Phi_\infty(z,\epsilon)=\left[\f{1}{4}F'(z)+\f{1}{4}F(z)-4z\right] \epsilon^{2/3}+\mathcal{O}(\epsilon).
\end{equation*}
From Eq.\eqref{eq:correction_to_scaling} we thus get the Riccati type ODE
\begin{equation}\label{eq:Fdiff}
F^\prime(z)=A z-B F(z)^2,
\end{equation}
where $A=16$ and $B=1$. In order to solve Eq.\eqref{eq:Fdiff}, we linearise it by using the ansatz
\BEN 
F(z)=b_0\f{f^\prime(b_1 z)}{f(b_1 z)},
\label{eq:ODE_linearising_ansatz}
\EEN
where 
\be 
b_0 = \left(\f{A}{B^2}\right)^{1/3} \text{ and } b_1 = \big(AB\big)^{1/3}.
\label{eq:scaling_parameters_DP}
\ee
This leads to the second order ODE
\be 
f''(z)-zf(z)=0,
\label{eq:Airy_ODE}
\ee
the general solution of which is given by
\be 
f(z) = \lambda_1 \op{Ai}(z) + \lambda_2 \op{Bi}(z),
\label{eq:general_solution_Airy_ODE}
\ee
where $\lambda_1,\lambda_2 \in \mathbb{R}$, the Airy function $\op{Ai}(z)$ is defined in Eq.\eqref{eq:def_Airy_function} and
\be 
\op{Bi}(z)=e^{-i\pi/6}\op{Ai}(ze^{-2i\pi/3})+ e^{i\pi/6}\op{Ai}(ze^{2i\pi/3}).
\ee	
Inserting the solution \eqref{eq:general_solution_Airy_ODE} into Eq.\eqref{eq:ODE_linearising_ansatz}, we obtain the general solution of Eq.\eqref{eq:Fdiff} as
\be 
F(z)=b_0\frac{(\lambda+1)\Ai'\left(b_1 z\right)+(\lambda-1)\Bi'\left(b_1 z\right)}{(\lambda+1)\Ai\left(b_1 z\right)+(\lambda-1)\Bi\left(b_1 z\right)},
\ee
where $\lambda \in \mathbb{R}$. It now follows from the asymptotic behaviour of $\op{Ai}(z)$, $\op{Bi}(z)$ and their derivatives (\href{http://dlmf.nist.gov/9.7}{$§$ 9.7 in }\cite{NIST}) that the only possibility to satisfy condition \eqref{eq:limit_behaviour_F} is to set $\lambda=1$. Thus, $F(z)$ has the form given in Eq.\eqref{eq:general_Airy_scaling_function}, and inserting the values $A=16$ and $B=1$ into Eq.\eqref{eq:scaling_parameters_DP}, we obtain 
\be 
b_0=b_1=2^{4/3}.
\ee

Note that this result is not given explicitly in \cite{Haug15}, since a different parametrisation was used in this reference. More precisely, the generating function $G(t,q)$ analysed there is related to the one discussed here by the relation $G(t,q)= G^{(\infty)}\big(\sqrt{t/q},\sqrt{q}\big)$, and therefore in \cite{Haug15}, $b_0=2$ and $b_1=4$. However, one verifies that both expressions are equivalent by substituting $t \to q t^2$ and $q\to q^2$ into the above result.

We now repeat the same analysis for Motzkin and Schr\"oder paths, for which the scaling behaviour has not yet been studied in the literature.

	\subsection{Motzkin paths ($\ell=1$)}
	
	Setting $\ell=1$ and $q=1$ in Eq.\eqref{eq:sol_gen_funeq_q=1}, we get the critical values for standard Motzkin paths as
	\begin{equation}
	G_c=\frac{1+2\sqrt{a}}{\sqrt{a}} \text{ and } t_c=\f{1}{1+2\sqrt{a}}.
	\end{equation}
	Analogous to the case of Dyck paths, we define $\Phi_1(z,\epsilon)$ from Eq.(\ref{eq:general_funeq_area-length}) as
	\begin{align*}
	\Phi_1(a,z,\epsilon)=&1-G_{sc}^{(1)}(a,z,\epsilon)+t(z,\epsilon)G_{sc}^{(1)}(a,z,\epsilon)+   \\ &+a(1-\epsilon)\,t(z,\epsilon)^2\,G_{sc}^{(1)}(a,z+t_c\epsilon^{1-\phi}-z\epsilon,\epsilon)\,G_{sc}^{(1)}(a,z,\epsilon).
	\end{align*}
	Again, assumption \eqref{eq:scaling_behaviour} implies that $\Phi_1(a,z,\epsilon)=\mathcal{O}(\epsilon^{3\theta})$ and requires the critical exponents to be $\theta=1/3$ and $\phi=2/3$. From the expansion
	\begin{align*}
	\Phi_1(a,z,\epsilon)=\left[a G_c t_c^4 F'(z)+at_c^3 F(z)^2-z(2 a G_c^2 t_c^2 + G_c )\right]   \epsilon^{2/3}+\mathcal{O}(\epsilon),
	\end{align*}
	we are then lead to the same ODE \eqref{eq:Fdiff} as for Dyck paths, with the coefficients now being
	\begin{equation}\label{eq:AB_Motzkin}
	A=\f{2 G_c}{t_c^2}+\frac{1}{at_c^3} \text{ and } B=\sqrt{a}.
	\end{equation}
	The final form of the scaling function is given by Eq.\eqref{eq:general_Airy_scaling_function} with
	\begin{equation}
	b_0=\left(\f{2\sqrt{a}+1}{a^2 t_c^3}\right)^{1/3} \text{ and }	b_1=\sqrt{a}\,b_0.
	\end{equation}

\subsection{Schr\"oder paths ($\ell=2$)}
\label{section:schroederpaths}

For Schr\"oder paths, the critical values are given by
\begin{equation}
t_c=\sqrt{1+2\,a-2\sqrt{a(a+1)}} \text{ and } G_c=\f1{\sqrt at_c}.
\end{equation}
As for Dyck and Motzkin paths, we define
	\begin{align*}
		\Phi_2(a,z,\epsilon)=&1-G_{sc}^{(2)}(a,z,\epsilon)+t(z,\epsilon)^2\,G_{sc}^{(2)}(a,z,\epsilon)+\\&+a(1-\epsilon)\,t(z,\epsilon)^2\,G_{sc}^{(2)}(a,z+t_c\,\epsilon^{1-\phi}-z\epsilon,\epsilon)\,G_{sc}^{(2)}(a,z,\epsilon),
	\end{align*}
	and assumption \eqref{eq:scaling_behaviour} determines $\theta=1/3$ and $\phi=2/3$. Expanding $\Phi_2(a,z,\epsilon)$ in $\epsilon$ gives with these critical exponents and the above values for $t_c$ and $G_c$,
	\begin{align*}
		\Phi_2(a,z,\epsilon)=\left[ a G_c t_c^4 F'(z)+a t_c^3F^2(z)-z\left(2a G_c^2t_c^2 +2 G_c t_c\right)\right]\epsilon^{2/3}+\mathcal{O}(\epsilon),
	\end{align*}
 which again leads to Eq.\eqref{eq:Fdiff}, where the coefficients are now
	\begin{equation}\label{eq:AB_Schroeder}
			A=\f{2 G_c}{t_c^2} + \f{2}{a t_c^{2}} \text{ and }
			B=\sqrt{a}.
	\end{equation}
 	Thus, also for Schr\"oder paths, the scaling function is given by Eq.\eqref{eq:general_Airy_scaling_function}, with %
	\begin{equation}
	\label{eq:lambda_mu_schroeder}
	b_0=\left(\f{2\sqrt{a}+2\,t_c}{a^2 t_c^3}\right)^{1/3} \text{ and }
	b_1=\sqrt{a}\,b_0.
	\end{equation}
	
\noi In the next section we now generalise the results obtained so far to general $\ell$.
		
		\subsection{The case of general $\ell$}
		
	Now we assume $\ell$ to be any positive integer. In this general case, it is not possible to give an expression for the critical value $t_c$ as a function of $a$.
			
		\noi As in the special cases, we define
		%
		\begin{align}
				\Phi_\ell(a,z,\epsilon)=&1-G_{sc}^{(\ell)}\left(a,z,\epsilon\right)+t(z,\epsilon)^\ell G_{sc}^{(\ell)}\left(a,z,\epsilon\right)+   \nonumber \\ 
		&+a(1-\epsilon) t(z,\epsilon)^2\,G_{sc}^{(\ell)}\left(a,z+t_c\epsilon^{1-\phi}-z\epsilon,\epsilon\right) G_{sc}^{(\ell)}\left(a,z,\epsilon\right),
\label{eq:Phi_ell}		
		\end{align}
		and from the assumption that $\Phi_\ell(a,z,\epsilon)=\Ord(\epsilon^{3\theta})$ one obtains $\theta=1/3$ and $\phi=2/3$. Expanding the RHS of Eq.\eqref{eq:Phi_ell} in $\epsilon$ we get
		\begin{align*}
		\Phi_\ell(s,z,\epsilon)=\left[ a G_c t_c^4 F'(z)+at_c^3F(z)^2-(2 a G_c^2 t_c^2+\ell\,G_c t_c^{\ell-1})z\right]   \epsilon^{2/3}+\mathcal{O}(\epsilon),
		\end{align*}
		which leads to Eq.\eqref{eq:Fdiff} with
		\be 
		A = \f{2 G_c}{t_c^2} + \f{\ell\,t_c^{\ell-4}}{a} \text{ and }
		B = \sqrt{a}.
		\ee	
		The solution of this equation is given in Eq.\eqref{eq:general_Airy_scaling_function} with parameters

	\begin{equation}
	b_0=\left(\f{2\sqrt{a}+\ell\,t_c^{\ell-1}}{a^2 t_c^3}\right)^{1/3} \text{ and }
	b_1=\sqrt{a}\,b_0.
	\end{equation}
	
In the next section we are going give a quick derivation of the solution for the generating function of Schr\"oder paths which is alternative to the one given in \cite{Oppenheim02}. This will make it possible to analyse the scaling behaviour of $G^{(2)}(a,t,q)$ rigorously by means of the method of steepest descents, and compare the rigorous result with the one obtained heuristically in section \ref{section:schroederpaths}.
	
\section{Exact solution for Schr\"oder paths ($\ell=2$)}
\label{section:solution_schroederpaths}


The exact solution for the area-width generating function of Schr\"oder paths ($\ell=2$) has been derived in Eq.(4.41) of \cite{Oppenheim02} by using the Enumerating Combinatorial Objects (ECO) method, and in this reference also a refined model with additional weights corresponding to the number of contacts of the walker with the line $y=0$ was considered. Here we give an alternative derivation of this result.
 
Substituting $\ell=2$ into Eq.\eqref{eq:general_funeq_area-length}, we obtain the functional equation
\be 
1 - t^2 G^{(2)}(a,t,q) + a q t^2 G^{(2)}(a,qt,q)G^{(2)}(a,t,q)=0
\label{eq:functional_eq_schroederpaths}
\ee
for the area-width generating function of Schr\"oder paths. Inserting the ansatz
\be 
G^{(2)}(a,t,q) = \f{H(a,q t^2,q)}{H(a,t^2,q)}
\label{eq:sol_schroederpaths}
\ee
into Eq.\eqref{eq:functional_eq_schroederpaths} for $\ell=2$, we get the linearised functional equation
\be 
a q t^2H(q^2 t^2)+(t^2-1) H(q t^2)+H(t^2) = 0,
\label{eq:linearized_fun_eq}
\ee
where we have abbreviated $H(t^2)\equiv H(a,t^2,q)$ for convenience. The fact that $t$ only appears quadratic in Eq.\eqref{eq:linearized_fun_eq} makes this equation easier to solve than the linearised functional equation for standard Motzkin paths. Namely, Eq.\eqref{eq:linearized_fun_eq} is solved by a basic hypergeometric series \cite{Gasper90}, defined as
\be 
H(t) = \sum_{n=0}^\infty \f{(-aq;q)_n}{(q;q)_n} q^{{n\choose 2}}(-t)^{n} = {_1\phi_1}\left(\BA{c} -a q \\ 0 \EA ; q,t \right),
\label{eq:1_phi_1}
\ee
where the $q$-Pochhammer symbol is given by (\href{http://dlmf.nist.gov/17.2#i}{$§$ 17.2} in \cite{NIST})
\be 
(z;q)_{n} = (1-z)\cd(1-q\,z)\cdots (1-q^{n-1}\,z).
\ee
This result can be verified by straightforward substitution. For $a=1$, $G^{(2)}(a,t,q)$ generates a $q$-deformation of the large Schr\"oder numbers as defined in \cite{Bonin93}.\\

\noi As an aside, we note that 
\be 
G^{(2)}\left(\f{1}{(h-1)\,q},t\,\sqrt{q(h-1)},q\right) = G_{p}(h,t,q),
\label{eq:peak_gf_DP}
\ee
where $G_p(h,t,q)$ is the generating function of Dyck paths, with weights $h,t$ and $q$ associated to the number of peaks, width and area, respectively. Here, a peak is called any up-step followed by a down-step. To verify Eq.\eqref{eq:peak_gf_DP}, it suffices to see that via a factorization argument similar to the one used in the derivation of Eq.\eqref{eq:general_functional_eq}, it follows that $G_{p}(h,t,q)$ satisfies the functional equation
\be 
1 +\big((h-1) qt^2-1\big) G_{p}(t) + qt^2 G_{p}(qt)G_{p}(t) = 0,
\ee
where $G_p(t) \equiv G_p(h,t,q)$. From Eq.\eqref{eq:general_functional_eq} it follows that the LHS of Eq.\eqref{eq:peak_gf_DP} satisfies the same functional equation, and therefore both functions are identical. Since for a Dyck path, the number of valleys equals the number of peaks minus one, the weight of peaks corresponds physically to the bending rigidity of the vesicle membrane, as it is incorporated in more realistic models \cite{Seifert97}. \\

In order to validate our results from Section \ref{section:schroederpaths}, we will now analyse the scaling behaviour of $G^{(2)}(a,t,q)$ by carrying out rigorous saddle point analysis. The same technique has been applied before to area-perimeter weighted staircase polygons and area-width-weighted Dyck paths \cite{Prellberg95,Haug15}.\\

\section{Saddle point asymptotics for Schr\"oder paths ($\ell=2$)}
\label{section:saddlepoint_asymptotics}

We are now going to analyse the asymptotic behaviour of the area-width generating function of Schr\"oder paths $G^{(2)}(a,t,q)$ around the tri-critical point $(a,t,q)=(a,t_c,1)$ of the model by using the method of steepest descents, generalised to the case of two coalescing saddle points. To this purpose, we use the exact solution for $G^{(2)}(a,t,q)$, which has been derived in the last section. We will first derive the asymptotics of the functions $H(t^2)$ and $H(qt^2)$ defined in Eq.\eqref{eq:1_phi_1}, and then use Eq.\eqref{eq:sol_schroederpaths} to obtain the leading asymptotic behaviour of $G^{(2)}(a,t,q)$. The calculation is analogous to the one carried out for staircase polygons and Dyck paths in \cite{Prellberg95,Haug15}, therefore we will only outline the essential steps here. We will further assume that $a>0$.

The first step is to write the series $H(t^2)$ as a contour integral. Using the formula
\be 
\f{(-1)^{n+1}q^{{n\choose 2}}}{(q;q)_n(q;q)_\infty}=\op{Res}\left[(z;q)_\infty^{-1} ; z=q^{-n}\right],
\ee
valid for $n \in \mathbb{N}$, one obtains the expression
\be 
H(t^2) = \lim_{N\to \infty} \f{A}{2\pi i}\int_{C_N} \f{z^{-2 \log_q(t)}}{(-aq/z;q)_\infty (z;q)_\infty} dz
\label{eq:exact_integral_formula}
\ee
where $\log_q$ denotes the logarithm with base $q$, $A = (q;q)_\infty(-aq;q)_\infty$ and the contour $C_N$ surrounds exactly the $N$ leftmost zeros of $(z;q)_\infty$. By estimating the integral in Eq.\eqref{eq:exact_integral_formula}, one then shows that in the limit $N\to \infty$, the contour $C_N$ can be exchanged by a straight line running from $\rho - i\infty$ to $\rho + i \infty$, where $0<\rho<1$. The two $q$-products in the denominator of the integrand in Eq.\eqref{eq:exact_integral_formula} can be estimated by using the formula \cite{Prellberg95}
\be 
\ln(z;q)_\infty = \f{1}{\ln(q)}\op{Li}_2(z)+\f12 \ln(1-z) + \ln(q) R(z,q),
\label{eq:asymptotics_q_product}
\ee
where $\op{Li}_2(z)$ is the Euler dilogarithm (\href{http://dlmf.nist.gov/25.12#E1}{Eq. 25.12.1} in \cite{NIST}) and  the remainder is sufficiently bounded in the relevant region of the $\mathbb{C}$ plane if $q\to 1^-$. Inserting Eq.\eqref{eq:asymptotics_q_product} into Eq.\eqref{eq:exact_integral_formula}, we obtain the asymptotic expression
\be 
H(t^2) = \f{A}{2\pi i}\int_{\rho-i\infty}^{\rho+i\infty}\exp\left(\f{1}{\epsilon}f(z)\right)\,g(z)\big(1+\mathcal{O}(\epsilon)\big)\,dz,
\label{eq:asympt_int_formula}
\ee
as $q\to 1^-$, where $\epsilon=-\ln(q)$,
\be 
f(z) &=& 2 \ln(t)\ln(z)+\Li_2(z)+\Li_2(-a/z),\text{ and}\\
g(z) &=& \sqrt{\f{z}{(1-z)(z+a)}}.
\ee
The function $f(z)$ has the two saddle points
\be 
z_{1}=\f{1}{2}\left(1-t^2-\sqrt {d}\right)\quad;\quad z_{2}=\f{1}{2}\left(1-t^2+\sqrt {d}\right)
\label{eq:saddlepoints_f}
\ee
where $d=(1-t^2)^2-4\,a\,t^2$. For $d=0$, the saddles coalesce in $z_m=(1-t^2)/2$. It is now possible to apply the method of steepest descents to the integral on the RHS of Eq.\eqref{eq:asympt_int_formula} (see e.g. \cite{Flajolet09_8} for a general introduction). 

\noindent From Theorem 1 in \cite{Chester57} it follows that there exists a transformation $u\mapsto z(u)$ which is regular in a domain containing $z_1$ and $z_2$ if $d$ is sufficiently close to zero, such that
\be 
f(z) = \f{1}{3}\,u^3-\alpha\,u+\beta = p(u).
\label{eq:trafo}
\ee
The polynomial $p(u)$ has saddle points $u_{1,2}=\pm \sqrt{\alpha}$. Since the transformation is regular, it necessarily maps these saddle points onto the two saddle points of $f(z)$, given in Eq.\eqref{eq:saddlepoints_f}. From Eq.\eqref{eq:trafo} we therefore obtain that 
\be 
\alpha=\left(\f34\big(f(z_1)-f(z_2)\big)\right)^{2/3}\text{ and } \beta = \f12\big(f(z_1)+f(z_2)\big).
\ee
Applying the transformation defined by Eq.\eqref{eq:trafo}, Eq.\eqref{eq:asympt_int_formula} can be rewritten as
\be 
H(t^2) = \f{A}{2\pi\,i}\int_{c_-\infty}^{c_+\infty}\exp\left(\f{1}{\epsilon}p(u)\right)\,S(u)\big(1+\mathcal{O}(\epsilon)\big)\,du,
\label{eq:asympt_int_formula_transformed}
\ee
with $S(u) = g(z(u))z'(u)$ and $c_\pm = \exp(\pm\,i\,\pi/3)$. In order to obtain the leading term of the asymptotic expansion of $H(t^2)$, we write
\be 
S(u) = p^{(0)}+u\,q^{(0)}+(u^2-\alpha)S_1(u),
\label{eq:g_times_Jabobian_expansion}
\ee
where $S_1(u)$ is a regular function of $u$, and the coefficients $p^{(0)}$ and $q^{(0)}$ can be determined by using that $S(u_{1,2})= p^{(0)}\pm\sqrt{\alpha}\,q^{(0)}$ and  inserting the saddle point values into the second derivative of Eq.\eqref{eq:trafo} to obtain $z'(u_{1,2})$. This gives the expressions
\be 
p^{(0)}&=&\sqrt{\f{\sqrt{\alpha}}{2}}\left(\f{g(z_2)}{\sqrt{f^{\prime\prime}(z_2)}}+\f{g(z_1)}{\sqrt{-f^{\prime\prime}(z_1)}}\right),\label{eq:p0}\\
q^{(0)}&=&\sqrt{\f{1}{2\sqrt{\alpha}}}\left(\f{g(z_2)}{\sqrt{f^{\prime\prime}(z_2)}}-\f{g(z_1)}{\sqrt{-f^{\prime\prime}(z_1)}}\right).
\label{eq:q0}
\ee
Inserting Eq.\eqref{eq:g_times_Jabobian_expansion} into Eq.\eqref{eq:asympt_int_formula_transformed} and using the coefficients given in Eq.(\ref{eq:p0}-\ref{eq:q0}), we arrive for $q\to 1^-$ at the asymptotic expression
\be 
H(t^2) \sim A\,\exp\left(\f{\beta}{\epsilon}\right)\left(\epsilon^{1/3} p^{(0)} \Ai\left(\f{\alpha}{\epsilon^{2/3}}\right)-\epsilon^{2/3}\,q^{(0)} \Ai^\prime\left(\f{\alpha}{\epsilon^{2/3}}\right)\right).
\label{eq:phi_expansion}
\ee

The asymptotic expression of $H(q^2 t)$ has the same form as Eq.\eqref{eq:phi_expansion}, with $p^{(0)}$ and $q^{(0)}$ replaced by the coefficients $p^{(1)}$ and $q^{(1)}$, which are obtained by replacing $g(z)$ by $h(z) = g(z)/z$ in Eqs.(\ref{eq:p0}-\ref{eq:q0}). With this we arrive at the following result. For $q \to 1^-$,
\be 
G^{(2)}(a,t,q) \sim \f{p^{(1)} \Ai(\alpha \epsilon^{-2/3})-q^{(1)} \Ai^\prime(\alpha \epsilon^{-2/3}) \epsilon^{1/3}}{p^{(0)} \Ai(\alpha \epsilon^{-2/3})-q^{(0)} \Ai^\prime(\alpha \epsilon^{-2/3}) \epsilon^{1/3}},
\label{eq:gen_fun_schroeder_uniform_as}
\ee
where $\epsilon=-\ln(q)$. Note that this expession is uniform for a range of values of $t$ and $a$ including the critical point $d=0$. In particular, setting $t=t_c-z\epsilon^{2/3}$, Eq.\eqref{eq:gen_fun_schroeder_uniform_as} gives for $\epsilon\to 0^+$,
\be 
\label{eq:scaling_schroeder_saddlepoint}
G^{(2)}(a,t,q) = \f{1}{z_m}\left(1+\left(\f{q^{(0)}}{p^{(0)}}-\f{q^{(1)}}{p^{(1)}}\right)\f{\Ai^\prime(\alpha \epsilon^{-2/3})}{\Ai(\alpha  \epsilon^{-2/3})}\,\epsilon^{1/3}+\Ord\left(\epsilon^{2/3}\right)\right).
\ee
Expanding the coefficients $\left(q^{(0)}/p^{(0)}-q^{(1)}/p^{(1)}\right)$ and $\alpha$ up to linear order around the critical point, we obtain the coefficients given in Eq.(\ref{eq:lambda_mu_schroeder}), thereby confirming the validity of the heuristic scaling ansatz. 
	\begin{figure}[htbp]
			\centerline{\includegraphics[width=0.6\textwidth]{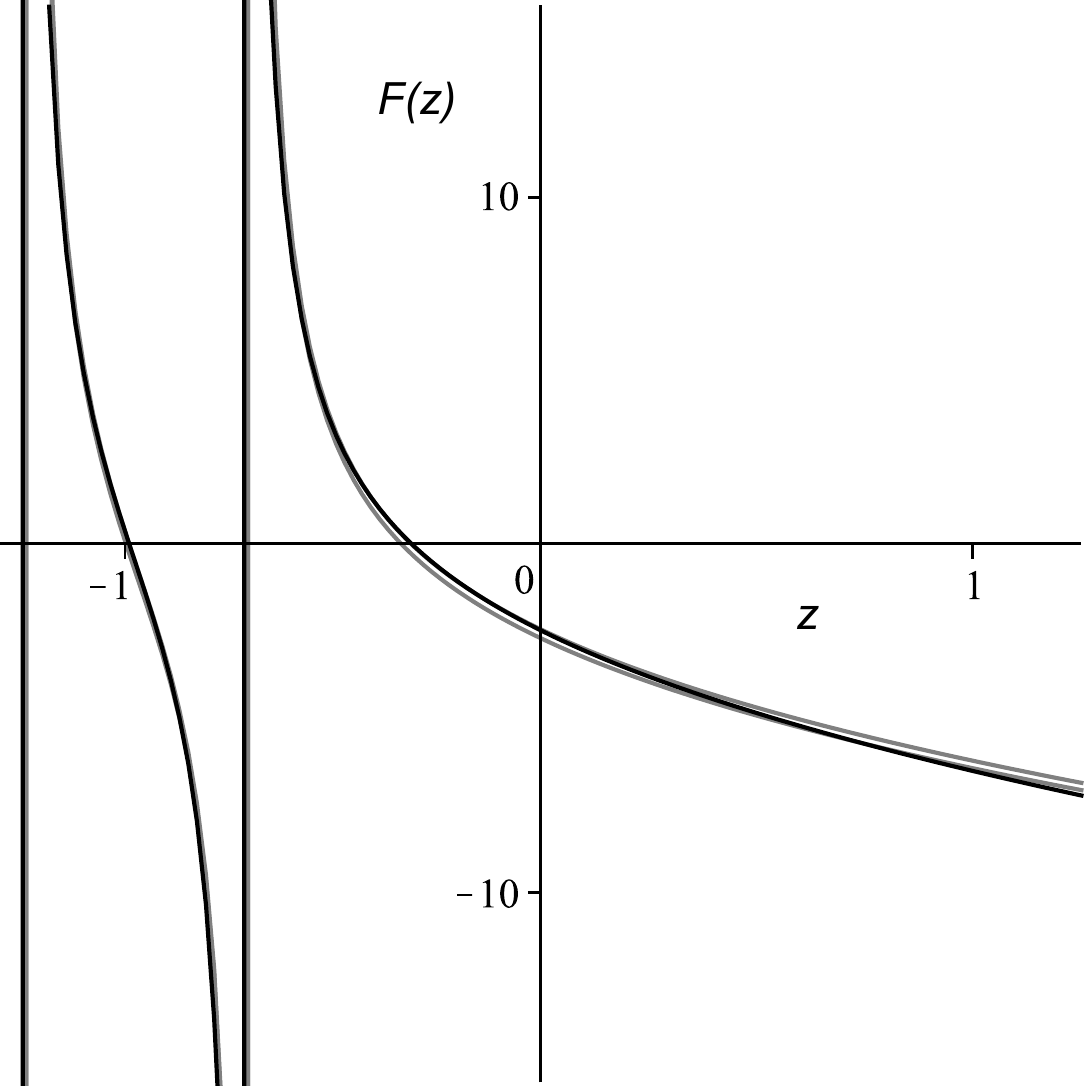}}
	\caption{Plot of the scaling function $F(z)$ given by Eq.\eqref{eq:general_Airy_scaling_function} with coefficients \eqref{eq:lambda_mu_schroeder} for $a=1$ (black) against the approximation of the scaling function obtained directly from the generating function $G^{(2)}(1,t,q)$ and fixed values 
	$\epsilon=10^{-3}$ and $10^{-4}$ (gray).}
		\label{fig:scalingfunctionplot}
	\end{figure}
	
Figure \ref{fig:scalingfunctionplot}~shows the remarkable agreement of scaling function and partition function asymptotics for $q$ close to one.

\section{Summary and Outlook}

We analysed the scaling behaviour of the generating function of area-weighted Dyck, Motzkin and Schr\"oder paths around the tri-critical point by using a heuristic ansatz and generalised this approach to $\ell$-Motzkin paths with arbitrary $\ell$. The scaling behaviour of Dyck paths had already been analysed in \cite{Haug15}. The result obtained is in agreement with the one derived in that reference. For Motzkin and Schr\"oder paths, no scaling function had been given yet in the literature. 

We gave a derivation of the area-width generating function of Schr\"oder paths alternative to the one in \cite{Oppenheim02}. The exact solution enabled us to obtain a rigorous result for the scaling behaviour of area-width weighted Schr\"oder paths by applying the saddle point method, thereby confirming the result obtained via the heuristic approach.

The solution for the area-width generating functions of Schr\"oder paths and Dyck paths is possible because in both cases, the parameter $t$ only appears in one power in both corresponding functional equations. The generating function for Motzkin paths has been obtained in \cite{Owczarek09}, but the derivation was more involved and led to a much more complicated expression. Exact solutions for the area-width generating functions for $\ell$-Motzkin with $2<\ell<\infty$ are not known yet. This therefore presents an open problem. 

Another open question is whether there exists a rigorous method to obtain the scaling behaviour of a generating function directly from the functional equation it satisfies, without having to know the exact solution. So far, the only rigorous method known is by applying the method of steepest descents, which requires knowledge of the exact solution.

\section*{Acknowledgement}

GS thanks Queen Mary University of London for hospitality and acknowledges the financial support from the doctoral scholarship from the National Science Centre in Poland (grant no. UMO-2015/16/T/ST1/00528).


\begin{thebibliography}{10}

\bibitem{Alberts07}
B.~Alberts, ``Intracellular vesicular traffic,'' in {\em Molecular biology of
  the cell}, ch.~13, Garland Science, Taylor and Francis Group, 2007.

\bibitem{Guttmann12}
A.~J. Guttmann, ``{Self-avoiding Walks and Polygons - An Overview},'' {\em Asia
  Pacific Mathematics Newsletter}, vol.~2, 2012.

\bibitem{Leibler89}
S.~Leibler, R.~R.~P. Singh, and M.~E. Fisher, ``Thermodynamic behavior of
  two-dimensional vesicles,'' {\em Phys. Rev. Lett.}, vol.~59, pp.~1989--1992,
  Nov 1987.

\bibitem{Fisher91}
M.~E. Fisher, A.~J. Guttmann, and S.~G. Whittington, ``Two-dimensional lattice
  vesicles and polygons,'' {\em J. Phys. A: Math. Gen.}, vol.~24,
  pp.~3095--3106, 1991.

\bibitem{Lawrie84}
I.~D. Lawrie and S.~Sarbach, ``Theory of tricritical points,'' in {\em Phase
  Transitions and Critical Phenomena} (C.~Domb and J.~L. Lebowitz, eds.),
  vol.~9, ch.~1, pp.~2--161, Academic Press, 1984.

\bibitem{Richard01}
C.~Richard, A.~J. Guttmann, and I.~Jensen, ``{Scaling function and universal
  amplitude combinations for self-avoiding polygons},'' {\em J. Phys. A: Math.
  Gen.}, vol.~34, pp.~L495--501, 2001.

\bibitem{NIST}
``{\it NIST Digital Library of Mathematical Functions}.''
  http://dlmf.nist.gov/, Release 1.0.13 of 2016-09-16.
\newblock F.~W.~J. Olver, A.~B. {Olde Daalhuis}, D.~W. Lozier, B.~I. Schneider,
  R.~F. Boisvert, C.~W. Clark, B.~R. Miller and B.~V. Saunders, eds.

\bibitem{Richard09}
C.~Richard, ``Limit distributions and scaling functions,'' in {\em Polygons,
  Polyominoes and Polycubes} (A.~J. Guttmann, ed.), vol.~775 of {\em Lecture
  Notes in Physics}, pp.~247--299, Springer Netherlands, 2009.

\bibitem{Richard02}
C.~Richard, ``Scaling behaviour of two-dimensional polygon models,'' {\em
  Journal of Statistical Physics}, vol.~108, no.~3, pp.~459--493, 2002.

\bibitem{Prellberg95}
T.~Prellberg, ``Uniform $q$-series asymptotics for staircase polygons,'' {\em
  J. Phys. A: Math. Gen.}, vol.~28, pp.~1289--1304, 1995.

\bibitem{Haug15}
N.~Haug and T.~Prellberg, ``{Uniform asymptotics of area-weighted Dyck
  paths},'' {\em J. Math. Phys.}, vol.~56, p.~043301, 2015.

\bibitem{Brak90}
R.~Brak and A.~J. Guttmann, ``Exact solution of the staircase and row-convex
  polygon perimeter and area generating function,'' {\em Journal of Physics A:
  Mathematical and General}, vol.~23, no.~20, p.~4581, 1990.

\bibitem{Flajolet80}
P.~Flajolet, ``{Combinatorial Aspects of Continued Fractions},'' {\em Discrete
  Mathematics}, vol.~32, pp.~125--161, 1980.

\bibitem{Krattenthaler15}
C.~Krattenthaler, ``Lattice path enumeration,'' in {\em Handbook of Enumerative
  Combinatorics} (M.~B\'ona, ed.), ch.~10, pp.~589--678, Boca Ranton: CRC
  Press, 2015.

\bibitem{Owczarek09}
A.~L. Owczarek and T.~Prellberg, ``{Exact solution of the discrete
  (1+1)-dimensional RSOS model with field and surface interactions},'' {\em
  Journal of Physics A: Mathematical and Theoretical}, vol.~42, no.~49,
  p.~495003, 2009.

\bibitem{Oppenheim02}
A.~C. Oppenheim, R.~Brak, and A.~L. Owczarek, ``Anisotropic step, surface
  contact, and area weighted directed walks on the triangular lattice,'' {\em
  Int. J. Mod. Phys. B}, vol.~16, no.~9, pp.~1269--1299, 2002.

\bibitem{Pergola02}
E.~Pergola, R.~Pinzani, S.~Rinaldi, and R.~A. Sulanke, ``A bijective approach
  to the area of generalized {M}otzkin paths,'' {\em Advances in Applied
  Mathematics}, vol.~28, no.~3, pp.~580 -- 591, 2002.

\bibitem{Haglund15}
J.~Haglund, ``Catalan paths and $q,t$-enumeration,'' in {\em Handbook of
  Enumerative Combinatorics} (M.~B\'ona, ed.), ch.~11, pp.~679--752, Boca
  Ranton: CRC Press, 2015.

\bibitem{wolfram_schroeder}
E.~W. Weisstein, ``{Schr\"oder Number}.''
  \url{http://mathworld.wolfram.com/SchroederNumber.html}, 2016.
\newblock From MathWorld--A Wolfram Web Resource.

\bibitem{wolfram_motzkin}
E.~W. Weisstein, ``Motzkin {N}umber.''
  \url{http://mathworld.wolfram.com/MotzkinNumber.html}.
\newblock From MathWorld--A Wolfram Web Resource.

\bibitem{OEIS}
{OEIS Foundation Inc.}, ``{The On-Line Encyclopedia of Integer Sequences}.''
  \url{http://oeis.org}, 2016.

\bibitem{Aigner07}
M.~Aigner, {\em {A Course in Enumeration}}.
\newblock Berlin Heidelberg: Springer--Verlag, 2007.

\bibitem{Gasper90}
G.~Gasper and M.~Rahman, {\em Basic Hypergeometric Series}, vol.~96 of {\em
  Encyclopedia of Mathematics and its Applications}.
\newblock Cambridge University Press, 1990.

\bibitem{Bonin93}
J.~Bonin, L.~Shapiro, and R.~Simion, ``Some $q$-analogues of the {S}chr\"oder
  numbers arising from combinatorial statistics on lattice paths,'' {\em J.
  Stat. Plan. Inference}, vol.~34, pp.~235--274, 1993.

\bibitem{Seifert97}
U.~Seifert, ``{Configurations of fluid membranes and vesicles},'' {\em Advances
  in Physics}, vol.~46, pp.~13--137, 1997.

\bibitem{Flajolet09_8}
P.~Flajolet and R.~Sedgewick, ``Saddle point asymptotics,'' in {\em {Analytic
  Combinatorics}}, ch.~8, Cambridge University Press, 2009.

\bibitem{Chester57}
C.~Chester, B.~Friedman, and F.~Ursell, ``An extension of the method of
  steepest descents,'' {\em Math. Proc. Cambridge Philos. Soc.}, vol.~53,
  pp.~599--611, 1957.

\end{thebibliography}

\end{document}